\documentstyle[11pt]{article}
\begin{document}

\begin{center}

{\LARGE\bf  Summary of Spin Physics Parallel Sessions}

\vskip 0.2in

{\large Jianwei Qiu$^*$ and Matthias Grosse Perdekamp$^\dagger$}

\end{center}

%\vskip 0.1in

\begin{quotation}

{\footnotesize \it 
\begin{center}
$^*$Department of Physics and Astronomy, Iowa State University, 
           Ames,~IA~50011

$^\dagger$Department of Physics, University of Illinois Urbana
	   Champaign, Urbana,~IL~61801 and RIKEN BNL Research Center, 
Upton,~NY~11973
\end{center} 
}
\end{quotation}

\begin{quote}
{\bf Abstract.}\ \
We summarize the activities in the spin physics parallel sessions of the
$8^{\rm th}$ conference on intersections between particle and nuclear
physics.
\end{quote}

\bigskip
\centerline{\large\bf Introduction}
\medskip

For spin physics in this conference, we had five parallel sessions and
over 500 minutes of presentations.  We had twenty-six scheduled talks 
%with twenty-four of them were presented (fourteen in experiment, two
with twenty-four of them presented at the conference (fourteen in
experiment, two 
in machine and instrumentation, and eight in theory). The talks
reported the activities at almost all major high energy spin
experiments around the world and covered a wild range 
of recent theoretical developments in this field.  The spin physics
parallel sessions had overwhelming participation and were full of 
exciting discussions. 

\bigskip
\centerline{\large\bf Polarized parton distributions}
\medskip

The determination of polarized quark distributions, $\Delta q(x,\mu)$, and 
the gluon distribution, $\Delta g(x,\mu)$, is essential for 
testing QCD perturbation theory in its spin sector as well as for
searching answers to the question on how the nucleon's spin is distributed
among its constituents.  
 
In the framework of QCD, the spin of a nucleon can be expressed as an
expectation value of the QCD angular momentum operator in the nucleon
state \cite{Filippone:2001ux},
\begin{equation}
\frac{1}{2} = \frac{1}{2}\, \Delta\Sigma(\mu) + \Delta g(\mu) 
            + L_q(\mu) + L_g(\mu)\, ,
\label{sumrule}
\end{equation}
where $\Delta\Sigma(\mu) \equiv \sum_q 
\int_0^1 dx [\Delta q(x,\mu)+\Delta \bar{q}(x,\mu)]$ and 
$\Delta g(\mu)=\int_0^1 dx \Delta g(x,\mu)$ are the total quark and
gluon helicity, respectively; and $L_q(\mu)$ and $L_g(\mu)$ are quark
and gluon orbital angular momentum (OAM), respectively.  The scale
$\mu$ indicates the momentum scale at which these quantities are
measured.  Combining all polarized deep inelastic scattering (DIS)
measurements, the quark helicity contribution to the proton spin is
found to be about
$\Delta\Sigma \sim 0.2$ \cite{Filippone:2001ux}, which is much smaller
than the unity expected from the naive quark model.  In order to test
the sum rule in Eq.~(\ref{sumrule}), we need to measure the individual
pieces on the right-hand-side.

The polarized parton distributions (pPDFs), $\Delta q(x,\mu)$ and 
$\Delta g(x,\mu)$ are nonperturbative but universal quantities, and
are not direct physical observables.  It is the QCD 
factorization theorem that connects them to polarized hadronic cross
%sections via perturbatively calculated partonic hard parts.  The
sections via perturbatively calculated partonic hard scattering processes. The
pPDFs can be extracted from measurements of polarized cross sections
with longitudinally polarized protons. 

Until the advent of the RHIC spin program, high energy collisions with
longitudinally polarized protons only took place in lepton-hadron DIS.
By measuring the double longitudinal spin asymmetry of inclusive DIS
cross sections, $A_{LL}$, one can extract the structure function,
$g_1(x,Q^2)$, of the polarized proton,  
\begin{equation}
g_1^p(x,Q^2) = \frac{1}{2}\left[ 
 \sum_{q,\bar{q}} \Delta C_q(x,\alpha_s) \otimes \Delta q(x,Q^2) 
+\Delta C_g(x,\alpha_s) \otimes \Delta g(x,Q^2) \right]
\label{g1-nlo} 
\end{equation} 
where the symbol $\otimes$ represents the convolution over the parton's
momentum fraction and the coefficient functions, $\Delta C_i$ with
$i=q,\bar{q},g$, are perturbatively calculable.  By fitting all
available data on the $g_1$ structure function for proton and neutron at
different values of Bjorken $x$ and scale $Q$ and using the DGLAP
evolution equation to control the $Q$-dependence of pPDFs, good
constraints on polarized valence quark distributions have been
obtained \cite{Wendtland,Hirai:qk}.  It was also found that the magnitude
of polarized sea distributions are much smaller than that of polarized
valence quark distributions and uncertainties on sea distributions are
much larger.    

The HERMES collaboration improved the separation of quark flavors and
extracted the sea and strange quark distributions by the measurement of 
semi-inclusive DIS (SIDIS) production of $\pi^{\pm}$ and $K^{\pm}$.  
With the help of existing quark-to-hadron fragmentation functions, they
reported the first determination of leading order (LO) $\Delta\bar{u}$,
$\Delta\bar{d}$, and $\Delta\bar{s}$ with the assumption of 
$[\Delta s/s](x) = [\Delta\bar{s}/\bar{s}](x)$ \cite{Wendtland}.  
The HERMES collaboration found no significant breaking of the flavor
symmetry in the light sea and no indication of a negative strange sea
contribution in fits to DIS data \cite{Wendtland}.  Kretzer in his
talk emphasized the importance of good knowledge of the relevant
fragmentation functions for the interpretation of these SIDIS
measurements \cite{Kretzer}. The HERMES collaboration also reports the
effort to explore the use of SIDIS $\Lambda$ production 
for improved sensitivity on $\Delta s$ \cite{Makins}.   

With the high luminosity at the Jefferson Laboratory (JLab), the double spin
asymmetry $A_{LL}$ in inclusive DIS was measured for the first time in
and near the resonance region, from which spin structure functions
were measured near $x=1$ \cite{Minehart}.

With a very successful 2002 run, the COMPASS collaboration recorded 1.2
fb$^{-1}$ DIS data with a longitudinally polarized target and 0.3 
fb$^{-1}$ DIS data with a transversely polarized target.
Reconstruction of events and data analysis are underway. The COMPASS
collaboration expects to have first physics results soon
\cite{Kabuss}.

The gluon helicity distribution, $\Delta g$, is a key quantity in our
understanding of the proton spin.  Unfortunately, because gluon
contributions only enter the $g_1$ structure function at
next-to-leading-order (NLO) order and indirectly via the 
DGLAP evolution, only limited information on $\Delta g$ has been obtained from
inclusive DIS measurements.  The Asymmetry Analysis Collaboration (AAC)
reports that global fits to all existing inclusive DIS data do not
provide good constraints on $\Delta g(x,\mu)$ and do not even
determine the sign 
\cite{Hirai:qk}.  With the fact that most hard processes at hadron
colliders are dominated by gluon initiated subprocesses, the RHIC spin
program will provide promising new information on $\Delta g$.

With the proton spin transversely polarized with respect to the
collision axis, a novel helicity flip {\it chiral-odd} twist-2 quark
distribution -- known as transversity distribution, $\delta q(x,\mu)$, is
theoretically allowed \cite{Ralston:79}. However, there is no leading
twist gluon transversity because it would require two units of helicity
flip.  Since perturbative hard processes conserve helicity, chiral-odd
distributions must appear in pairs.  That is, transversity can never
be measured in polarized inclusive DIS cross sections.

Although transversity can be in principle extracted from the
measurement of double transverse spin asymmetries, $A_{TT} \propto 
\delta q(x) \otimes \delta q(x')$, in jet- or inclusive
hadron production in polarized hadronic collisions,
the asymmetries are often too small to be extracted experimentally because
of the dominance (or lack) of gluonic contribution to the unpolarized
(or polarized) cross sections \cite{Vogelsang}. 

Therefore, $\delta q(x,\mu)$, is better determined from observables
dominated by quark-initiated partonic subprocesses, like $A_{TT}$   
of Drell-Yan, which unfortunately suffers from the low rate at the
luminosities presently available at RHIC.  The single transverse spin
asymmetry $A_{UT}$ in SIDIS between a unpolarized lepton
beam and a transversely polarized target is proportional to
a combination of $\delta q(x) \otimes \Delta D(z,{\bf k}_T)$ with the
Collins' ${\bf k}_T$-dependent fragmentation function 
$\Delta D(z,{\bf k}_T)$ \cite{Collins:1992kk}.  
Transversity distributions could also be accessed by measuring
$A_{UT}$ in SIDIS in combination with interference fragmentation
functions \cite{Jaffe:1997hf}.  

\bigskip
\centerline{\large\bf Single spin asymmetries in SIDIS}
\medskip

Single spin asymmetries (SSA) in SIDIS between a unpolarized lepton
beam and a polarized target can be achieved via the Collins
\cite{Collins:1992kk} and Sivers \cite{Sivers:1989cc} mechanisms at
leading twist or the Qiu-Sterman mechanism \cite{Qiu:pp} at twist-3.
Because of Lorentz invariance of QCD, we need at least four vectors
including the spin vector to construct a physically observed SSA.  
For example, SSAs in SIDIS should be proportional to 
$\epsilon_{\mu\nu\alpha\beta}\, q^{\mu} P^{\nu} S^{\alpha}
p^{\beta}$ where $q$ is momentum of the virtual photon, $P$ ($S$)
are momentum (spin) of the polarized target, and $p$ is the momentum of 
the observed final-state hadron.  Consequently, SSAs have a unique
$\sin(\phi)$ dependence with $\phi$ being the angle between the plane
of the three four-vectors ($q, P, S$) and the plane of the four-vectors
($q, P, p$).  

Let $q$ and $P$ define the collision $z$-axis and $p_T$ be the transverse
momentum of the observed hadron in this frame, then the SSA should be roughly
proportional to the dimensionless coefficient: $p_T M /(p_T^2 + M^2)$
with the typical hadronic mass $M \ll Q$ \cite{Collins:1992kk}.  
Therefore, SSAs in general have two distinct regions of phase space:
leading twist SSA when $p_T\ll M$ and twist-3 SSA when $p_T\gg M$.
When $p_T$ is small, while $Q$ in SIDIS provides the hard scale required for
the validity of the leading twist approximation, SSAs should be
proportional to $p_T/M$; 
and the Collins and Sivers mechanisms lead to nonvanishing SSAs: 
$A_{UT} \propto \delta q(x) \otimes \Delta D(z,{\bf k}_T)
 + \Delta f(x,{\bf k}_T) \otimes D(z)$.
The size and sign of the asymmetry are controlled by 
the nonperturbative ${\bf k}_T$-dependent Collins' fragmentation
function $\Delta D(z,{\bf k}_T)$ and the ${\bf k}_T$-dependent Sivers'
distribution function $\Delta f(x,{\bf k}_T)$.  When $p_T$ is of order
of the hard 
scale $Q$, SSAs should be proportional to $M/p_T$, a typical twist-3
behavior; and Qiu and Sterman show that the SSA is proportional to new
twist-3 tri-parton correlation functions and the corresponding partonic
hard scattering pieces can be systematically calculated in pQCD \cite{Qiu:pp}.
In terms of final-state interaction between outgoing quark and target
spectator, Brodsky et al. \cite{Brodsky:2002cx} explicitly
demonstrated that the Sivers contribution to SSAs in SIDIS does not vanish. 

Both the HERMES collaboration \cite{Seidel} and the CLAS collaboration
\cite{Avakian} observed SSAs in $e+P(S)\rightarrow e'+\pi(p)+X$ when
the target proton spins are oriented either along or perpendicular to the
direction of incoming lepton, and verified that both $A_{UL}$ and
$A_{UT}$ are proportional to $\sin(\phi)$. $A_{UL}$ is mainly
sensitive to the Collins effect while $A_{UT}$ should include
contributions from both the Collins and Sivers effects.  

On the theory side, Gamberg \cite{Gamberg} reported explicit model
calculation of SSAs for both $A_{UL}$ and $A_{UT}$.  Afanasev
\cite{Afanasev} reported a calculation of a beam SSA.  Instead of
the target spin vector, the beam SSA is due to a polarized virtual photon
on a unpolarized target in SIDIS.  The beam SSA is suppressed by an extra
power of $1/Q$ compared to the SSA from target spin.  

Hasuko \cite{Hasuko} reported the status of a program to measure
spin dependent fragmentation functions at the
Belle experiment in Japan. 
By measuring the final-state hadron azimuthal
asymmetry and correlation, it will be possible to extract the Collins 
function and possibly other chiral-odd fragmentation functions. 

\bigskip
\centerline{\large\bf Orbital angular momentum and generalized parton
  distributions} 
\medskip

Asymptotically, the quarks, including both helicity and OAM, carry
about 52\% of total proton spin while the gluons carry the rest
\cite{Ji:1996ek}.  Since quark helicity contributes about 20\% of
the proton's spin, a significant quark contribution to the proton's spin
must come from 
its OAM \cite{Filippone:2001ux}.  Therefore, knowing the OAM
contribution of partons is 
crucial for understanding the decomposition of the nucleon's spin in
Eq.~(\ref{sumrule}).  

Generalized parton distributions (GPDs) share the same operators as
normal parton distributions, but are evaluated with a pair of nucleon
states of different momenta.  GPDs carry much more information than
what parton distributions can provide.  Belitsky \cite{Belitsky}
emphasized that parton orbital angular momentum is directly related
to GPDs.  He also pointed out that deep virtual Compton scattering
(DVCS), recently measured at HERMES, H1 and CLAS, is the cleanest
hadronic reaction that gives access to GPDs. The HERMES collaboration
reported a measurement of GPDs from SIDIS meson production \cite{Hasch}.

Ji et al. \cite{Ji} introduced a phase-space distribution in terms of
the GPDs to describe the probability to find a quark at a given
momentum and position.  Parton OAM plays an important role in
determining the phase-space distribution because of its connection 
to GPDs.  The phase-space distribution provides the possibility to
study spatial distributions of partons inside a nucleon.

\bigskip
\centerline{\large\bf RHIC spin program}
\medskip

The RHIC spin program had its first polarized proton-proton collision
two years ago. Two brief polarized proton runs in 2002 and 2003 have 
resulted in first physics results on transverse spin asymmetries and 
many important advances in accelerating high current polarized proton beams 
to high energies as well as proton beam polarimetry at high energies.
The main goal of the program is to measure polarized parton distributions,
in particular the polarized gluon distribution.
 However, the combination of available spin orientations and a broad
 range of experimental processes will also allow to probe the physics 
 beyond the leading twist QCD dynamics as well as beyond the Standard Model.  

The STAR Collaboration \cite{Igo} has measured the single transverse spin
asymmetry $A_N$ for pion production both for positive and negative 
rapidity $y$.  In the 2003 run, spin rotators were commissioned and  
for the first time, double spin asymmetries $A_{LL}$ were measured in a
hadron collider. The data analysis for $A_{LL}$ in jet production is
in progress and will provide first new information on the gluon helicity
distribution \cite{Igo}. With similar physics motivation, the PHENIX
collaboration  
is carrying out analysis of double spin asymmetries in inclusive
hadron production \cite{Field}. 

In order to improve the figure of merit for the asymmetry measurements,
MacKay \cite{MacKay} argued that it will be possible for RHIC to raise the
beam polarization from 35-40\% to 70\%. He discussed possible methods and
plans to improve the luminosity of the polarized beam, which is now
about a factor 30 
below the design value. Bravar \cite{Bravar} showed that elastic
$p^{\uparrow} C \rightarrow p^{\uparrow} C$ has been used successfully
as a fast polarimeter at RHIC.

%\subsection{Summary and outlook}

With three DIS machines (HERMES, Compass, JLab) running at different 
energies and RHIC spin program in a full swing, we will soon have more
precise data to cover a wide spectrum of spin physics. A new
experiment to test the factorization of SIDIS was just proposed  
to JLab \cite{Jiang}.  Spin physics is getting more and more exciting
now! 

%\begin{theacknowledgments}
This work is supported by the U.S. Department of Energy  
under Grant No. DE-FG02-87ER40371.
%\end{theacknowledgments}

%\bibliographystyle{aipproc}   % if natbib is available


\vskip -0.1in

{\footnotesize
\begin{thebibliography}{80}

%\cite{Filippone:2001ux}
\bibitem{Filippone:2001ux}
B.~W.~Filippone and X.~D.~Ji,
Adv.\ Nucl.\ Phys.\  {\bf 26}, 1 (2001), and references therein.

\bibitem{Wendtland}
J. Wendtland, HERMES Collaboration, in this proceedings.

%\cite{Hirai:qk}
\bibitem{Hirai:qk}
M.~Hirai {\it et al.}, AAC, 
Int.\ J.\ Mod.\ Phys.\ A {\bf 18}, 1203 (2003); and references
therein. 

\bibitem{Kretzer}
S. Kretzer, in this proceedings.

\bibitem{Makins}
N. Makins, HERMES Collaboration, in this proceedings.

\bibitem{Minehart}
R. Minehart, CLAS collaboration, in this proceedings.

\bibitem{Kabuss}
E. Kabuss, Compass Collaboration, in this proceedings.

\bibitem{Ralston:79}
J. Ralston and D.E. Soper, Nucl. Phys. {\bf B152}, 109 (1979).

\bibitem{Vogelsang}
W. Vogelsang, in this proceedings.

\bibitem{Collins:1992kk}
J.~C.~Collins,
Nucl.\ Phys.\ B {\bf 396}, 161 (1993).

\bibitem{Jaffe:1997hf}
R.~L.~Jaffe, X.~m.~Jin and J.~Tang,
Phys.\ Rev.\ Lett.\  {\bf 80}, 1166 (1998).

\bibitem{Sivers:1989cc}
D.~W.~Sivers,
Phys.\ Rev.\ D {\bf 41}, 83 (1990); {\it ibid.}
{\bf 43}, 261 (1991).

\bibitem{Qiu:pp}
J.~w.~Qiu and G.~Sterman,
Phys.\ Rev.\ Lett.\  {\bf 67}, 2264 (1991);
Nucl.\ Phys.\ B {\bf 378}, 52 (1992).

\bibitem{Brodsky:2002cx}
S.~J.~Brodsky, D.~S.~Hwang and I.~Schmidt,
Phys.\ Lett.\ B {\bf 530}, 99 (2002); D.~S.~Hwang, in this proceedings.

\bibitem{Seidel}
R. Seidel, HERMES Collaboration, in this proceedings.

\bibitem{Avakian}
H. Avakian, CLAS Collaboration, in this proceedings.

\bibitem{Gamberg}
L. Gamberg, in this proceedings.

\bibitem{Afanasev}
A. Afanasev, in this proceedings.

\bibitem{Hasuko}
K. Hasuko, Belle collaboration, in this proceedings.

%\cite{Ji:1996ek}
\bibitem{Ji:1996ek}
X.~D.~Ji,
Phys.\ Rev.\ Lett.\  {\bf 78}, 610 (1997).

\bibitem{Belitsky}
A.V. Belitsky, in this proceedings.

\bibitem{Hasch}
D. Hasch, HERMES Collaboration, in this proceedings.

\bibitem{Ji}
X. Ji, in this proceedings;  
A. V. Belitsky, X. Ji and F. Yuan, arXiv:hep-ph/0307383.

\bibitem{Igo}
G. Igo, STAR Collaboration, in this proceedings.

\bibitem{Field}
D. Field, PHENIX Collaboration, in this proceedings.

\bibitem{MacKay}
W. MacKay, in this proceedings.

\bibitem{Bravar}
A. Bravar, in this proceedings.

\bibitem{Jiang}
X. Jiang, in this proceedings.
\end{thebibliography}

}
\end{document}